\documentclass{article}
\usepackage{inputenc}
\usepackage[sort,compress,numbers]{natbib}
\usepackage{graphicx,xcolor}
\usepackage{subfigure}
\usepackage{amsmath}
\usepackage{amssymb}
\usepackage[left=20mm, right=20mm, top=20mm, bottom=20mm]{geometry}
\newcommand{\be}{\begin{equation}
\newcommand{\ee}{\end{equation}}}
\newcommand{\bea}{\begin{eqnarray}}
\newcommand{\eea}{\end{eqnarray}}
\newcommand{\nn}{\nonumber}

\begin{document}
\markboth{ Ahmed Al-Jamel}{}

%
%

%
%

\title{\textbf{Saturation effect in confined quantum systems with energy-dependent potentials} }

\author{Ohood Ali  AL-Sbaheen$^{1}$, Ahmed Al-Jamel$^{1}$, Mohamed Ghaleb Al-Masaeed$^{2,3}$ \\
$^1$Physics Department, Faculty of Science, Al al-Bayt University,\\ P.O. Box 130040, Mafraq 25113, Jordan\\$^2$Ministry of Education, Jordan\\ $^3$Ministry of Education and Higher Education, Qatar\\ohoodsbba996@gmail.com\\
aaljamel@aabu.edu.jo, aaljamel@gmail.com\\moh.almssaeed@gmail.com}

\maketitle

\begin{abstract}
In this paper, we study the saturation effect in the energy or mass spectra of three quantum models with energy-dependent potentials: the harmonic oscillator, the hydrogen atom, and the heavy quarkonia. We used the method proposed in \cite{garcia2009exactly}, which is based on studying various canonical point and gauge transformations applied to a function, $g(x)$, multiplied by a given differential equation of known solutions as special orthogonal functions, that convert it into a Schr{{\"o}}dinger-like equation. The first two models stem from implementing the method on the confluent hypergeometric differential of the well-known solutions ${}_1 F_1$, while the third model (heavy quarkonia) stems possibly from the hypergeometric differential of the well-known solutions ${}_2 F_1$. In particular, the heavy quarkonia mass spectra for both $c\bar c$ and $b\bar b$ are produced at different values of the saturation parameter $\lambda$ and compared with the available experimental data. It is found that these systems may exhibit saturation effect when the energy-dependent effect is included.

\end{abstract}
\section{Introduction}
In the quantum world, the quest for bound-state solutions to wave equations, whether relativistic or non-relativistic, is of prime importance to investigate their spectroscopic as well as other possible properties. Many papers have been written on this topic, using different methods and proposed phenomenological potential models. Examples of such potential models are Coulomb, Martin, Woods-Saxon, Eckart, trigonometric Rosen-Morse, logarithmic, harmonic, and Cornell potentials, each of which has properties that are applicable to use for the physical problem under study. For instance, the Cornell potential is an ideal model for studying heavy quarkonia because it takes into account the two unique features of the strong interaction, namely, confinement and asymptotic freedom.
One of the interesting properties that the confining potential model should take into consideration is the saturation effect, viz. the quenching of indefinite growth in the energy spectra. Otherwise, the energy spectrum increases with increasing the quantum number which then leads to unstable systems. The saturation effect could be achieved by considering energy-dependent potentials. The wave equations with such energy-dependent potentials are already familiar in physics. For instance, the relativistic description of a scalar particle in the presence of an external Coulomb field through the Klein–Gordon equation leads to a wave equation with an energy-dependent potential \cite{rizov1985relativistic}, the Pauli - Schr{\"o}dinger equation describing a particle in an external electromagnetic field \cite{pauli_zur_1927}, and in non-relativistic quantum mechanics, they arise from momentum-dependent interactions \cite{green1962velocity}. 

Formanek et al \cite{formanek2004wave} have investigated wave equations that involve potentials that depend on the energy of the system and a formal analysis to determine the conditions under which these wave equations can be treated as evolution equations of quantum theory. In Ref \cite{langueur2019dkp}, has been discussed a study on the DKP equation with energy-dependent potentials, including the calculation of normalization and continuity equations, the determination of eigenfunctions and eigenvalues, and the presentation of energy graphs. Also, have been generalized Schr{\"o}dinger equations with energy-dependent potentials through formalism and applications \cite{10.1063/1.5058145}. Besides, have been studied  of heavy quarkonium with energy-dependent Potential \cite{gupta2012study}.\\

Additionally, Ref \cite{doi:10.1142/S0217732319503073} used the Nikiforov-Uvarov method to solve the radial Schr{\"o}dinger equation for heavy quarkonia confined within a potential that accounts for conformal symmetry, and found that a small perturbing term in the potential achieves good agreement with experimental data. The formalism of quantum mechanics is extended to a variety of systems that have energy-dependent potentials and position-dependent masses, resulting in modifications to the scalar product and norm \cite{10.1063/1.5058145}. Furthermore, Point transformations were used to develop energy-dependent potentials, with instances of boundary-value problems with a discrete spectrum and closed-form solutions, and solutions in terms of exceptional orthogonal polynomials \cite{schulze-halberg_quantum_2017}.

In the situation of energy-dependent potentials for bound states, the properties of the wave equation were examined \cite{yekken2010energy}. Besides, Within the framework of the path integral approach, the normalization problem of energy-dependent potentials was analyzed and that was applied to the harmonic oscillator and the hydrogen atom (radial) \cite{benchikha2013energy}. Moreover, in Ref \cite{tilaver2021energy}, the focus was on how the energy-dependent barrier-type (rectangular, step, Hulthn, Woods–Saxon) characteristic affects the transmission and reflection coefficients of the Klein–Gordon and Dirac particles.\\

In ref \cite{lombard2007wave}, the wave equation with energy-dependent confining potentials admitting analytic solutions, namely the linear and harmonic potentials has been studied. It is found that for linear energy dependence, the energy spectrum exhibits a saturation effect. A toy model was also presented to produce the heavy quarkonia spectra. In Ref \cite{lombard2009many}, a system of $N$ bosons with mutual two-particle confining harmonic interactions with energy-dependent frequency has been investigated. Such a situation was found to produce noticeable properties of the many-body system such as saturation behavior.

In this work, we investigate the saturation effects in the energy or mass spectra of confined systems in different energy-dependent potential models. Rather than starting from the Schr{\"o}dinger equation and substituting a proposed potential, we follow a different approach developed in \cite{garcia2009exactly}.  The approach is based on finding the set of energy-dependent potentials for which the Schr{\"o}dinger equation permits solutions in terms of known orthogonal polynomials such as the hypergeometric functions, Hermit polynomials, and others. The method is based on studying various canonical point and gauge transformations applied to a function, $g(x)$, multiplied by a differential equation related to a known special function, that converts it into a Schr{\"o}dinger -like equation. We then search for the saturation effects in the energy spectra obtained for these potentials. As an application, we search for saturation in mass spectra of heavy quarkonia and compare our results with the available experimental results.

\section{Theoretical Background}
We briefly review here the method for constructing Schr{\"o}dinger equations with energy-dependent potentials that are exactly solvable; the full presentation is discussed in \cite{garcia2009exactly}. We consider the second-order differential equation with variable coefficients of the form \cite{garcia2009exactly}:
\be
\label{29}
\mathcal{L}_x y(x) = 0,
\ee
with the operator $\mathcal{L}_x$
\be
\label{210}
\mathcal{L}_x  = P(x)  \frac{d^2 }{dx^2} +Q(x)  \frac{d}{dx}+R(x),
\ee
such that for the coefficient functions $P(x), Q(x)$, and $R(x)$, the solution $y(x)$ is known. We then consider the influence of multiplying Eq. (\ref{29}) by an arbitrary global multiplicative factor $g(x)$, viz., 
\be
\label{211}
g(x)[\mathcal{L}_x y(x)] = 0,
\ee
 and using the change of variable 
  \be
\label{212}
x=\mathcal{F}(u).
\ee
Here $\mathcal{F} (u)$ is an unknown function for the present. Following the procedure and steps in \cite{garcia2009exactly}, $u$ is given as
\be
\label{215}
u=\pm \int^{\mathcal{F} (u)} \frac{d \mathcal{F} (u)}{\sqrt{g(\mathcal{F} (u)) P(\mathcal{F} (u))}} = \pm \int^{x} \frac{dx}{\sqrt{g(x) P(x)}}, 
\ee
in terms of the coordinate $(u)$ is
\be
\label{219}
\mathcal{L}_u y (\mathcal{F} (u)) =0 .
\ee
and using the similarity transformation:
\be
\label{220}
y(x) = y (\mathcal{F} (u)) = \phi (u) e^{-\int^u W(u) du},
\ee
where
\be
\label{218}
W(u)= \frac{g(\mathcal{F} (u)) Q(\mathcal{F} (u)) - \mathcal{F}^{''} (u)}{2\mathcal{F}^{'} (u)},
\ee

we obtain
 \be
\label{221}
\phi (u) = e^{\int^u W(u) du}  y (\mathcal{F} (u)).
\ee
Eq. (\ref{219}) can be transformed into the next equation without the first derivative
 \be
\label{222}
- \frac{d^2 \phi (u)}{du^2} + [v(u)-g(\mathcal{F} (u)) R(\mathcal{F} (u))] \phi (u) = 0,
\ee
where:
 \be
\label{223}
v(u) = W^2(u) + W^{'}(u).
\ee
It is fruition at this moment to note that Eq. (\ref{222}) is of the form of Schr{\"o}dinger equation with potential $v(u)$, whose solution is the known function $\phi(u)$ given in Eq. (\ref{221}). Thus, the introduction of the function $g(x)$ in Eq.(\ref{211}), modifies the potential in Eq. (\ref{222}), via Eq.(\ref{218}), which leads to a set of Schr{\"o}dinger equations from Eq.(\ref{29}). Every choice $g(x)$ generates a corresponding Schr{\"o}dinger equation. Writing 
 \be
\label{224}
\frac{\hbar^2}{2m} [v(u)-g(\mathcal{F} (u)) R(\mathcal{F} (u))] \equiv V(u)-E,
\ee
 then Eq.(\ref{223}) reduces to the Schr{\"o}dinger equation
 \be
\label{225}
-\frac{\hbar^2}{2m} \frac{d^2 \phi(u)}{du^2} + V(u) \phi(u)= E\phi(u).
\ee
 We discuss three examples of interest in physics.
 The harmonic oscillator and the hydrogen atom, which stem from the confluent hypergeometric equation, and the heavy quarkonia mass spectra which stems from The hypergeometric equation (not confluent).
  \section{Results and discussion}
 Here, we present three ideal quantum systems of interest: the harmonic oscillator, the hydrogen atom, and the heavy quarkonia. The saturation effect in the energy spectra of these systems with energy-dependent potential is investigated in terms of the energy parameters. Following \cite{garcia2009exactly} and Tables therein, the harmonic oscillator and the hydrogen atom eigen-solutions can be generated from the following confluent hypergeometric equation
\be
\label{31}
x y^{''} + (c-x) y^{'} -ay = 0,
\ee
which has first solution of the form:
 \be
 \label{32}
y(x)= {}_1 F_1\left(a; c;x\right), c \in \mathbb{Z^{-}},
 \ee
 where ${}_1 F_1 \left(a; c;x\right) $ is the confluent hypergeometric function, which becomes a polynomial if $a \in \mathbb{Z^{-}}$. Remember, we're searching for potentials and energies that can produce these solutions. We then implement the method described in Sec. 2 to Eq. (\ref{31}) with appropriate choices of $g(x)$.
 \subsection{The harmonic oscillator}
\begin{figure}[!htbp]
  \centering
  \subfigure [] {\includegraphics[scale=0.7]{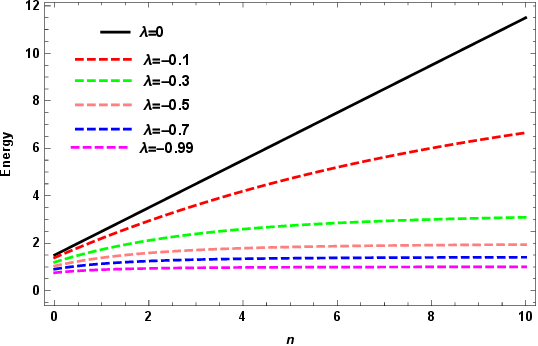}}\quad
  \subfigure []{\includegraphics[scale=0.7]{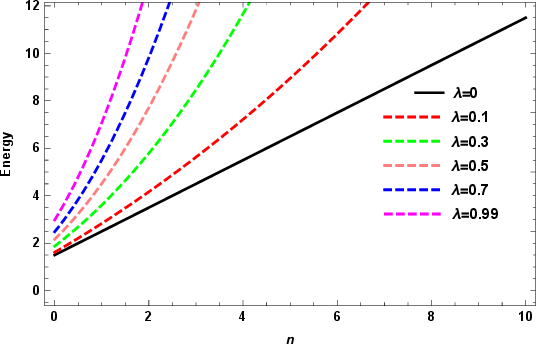}}
  \hfill
  \caption{ $E_n$ as function of $n$ for the linear E-dependent harmonic oscillator, with $q=1$ and different values of $\lambda$.}
	\label{H1}
\end{figure}

\begin{figure}[!htbp]
  \centering
  \subfigure [] {\includegraphics[scale=0.7]{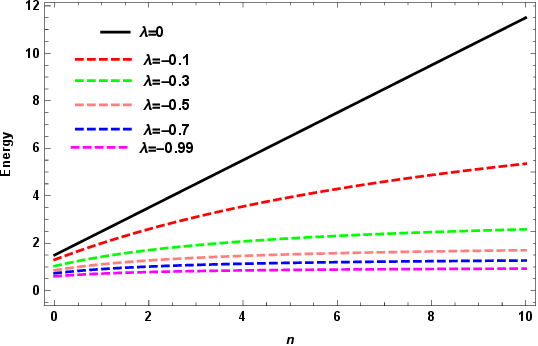}}\quad
  \subfigure []{\includegraphics[scale=0.7]{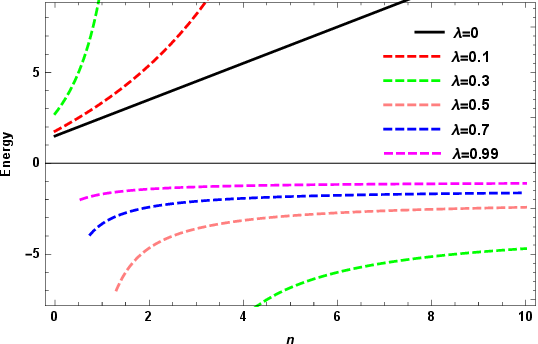}}
  \hfill
  \caption{ $E_n$ as function of $n$ for the linear E-dependent harmonic oscillator, with $q=2$ and different values of $\lambda$.}
	\label{H2}
\end{figure}

\begin{figure}[!htbp]
  \centering
  \subfigure [] {\includegraphics[scale=0.7]{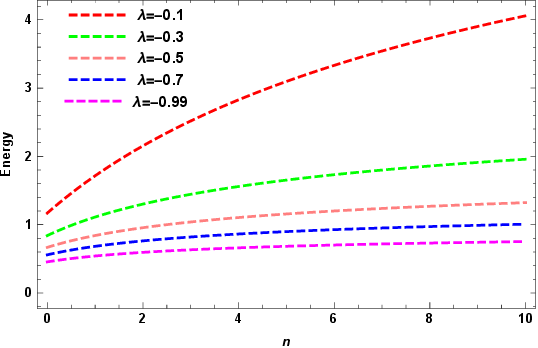}}\quad
  \subfigure []{\includegraphics[scale=0.7]{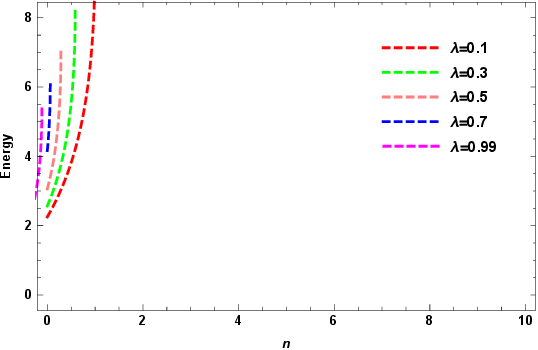}}
  \hfill
  \caption{ $E_n$ as function of $n$ for the linear E-dependent harmonic oscillator, with $q=4$ and different values of $\lambda$.}
	\label{H4}
\end{figure}

  We calculate the energy spectrum of the harmonic oscillator using the confluent hypergeometric equation. Energy spectrum $E_n$, and solutions $\phi(u)$ for several energy-dependent potentials $V (E_n;u)$ obtained from the method of chapter 2 when it is applied to the confluent hypergeometric equation of the harmonic oscillator:
Following \cite{garcia2009exactly}, the choice 
 \be
 \label{33}
 g(x)=k^2,
 \ee
where $k$ is a constant, generates the eigen-solutions for the harmonic oscillator. Using Eq.~(\ref{215}), and the coordinate change in Eq.~(\ref{212}), we have 
 \bea
 \nn
 u= \pm \int^x \frac{dx}{\sqrt{k^2 x}} = \pm \frac{2 \sqrt{x}}{k}.
 \eea
 Using the change of variable  $u=\pm \frac{2 \sqrt{x}}{k}$, yields 
 \be
 \label{34}
\mathcal{F} (u) =  \pm \frac{ u^2 k^2}{4}.
 \ee
 Then, with the help of Eq. (\ref{218}), we have
 \be
 \label{35}
 W(u) = \frac{2c-1}{2u} - \frac{u k^2}{4},
 \ee
 where $\mathcal{F}^{''} (u) = \frac{1}{2} k^2, Q(F(u))= c-\frac{u^2 k^2}{4}$. Substituting this result in Eq.~(\ref{223}), we obtain
 \be
 \label{36}
 v(u) = \frac{4c^2+3-8c}{4u^2} + \frac{u^2 k^4}{16} - \frac{ck^2}{2},
 \ee
 and using Eq.~(\ref{224}), we have 
 \be
 \label{37}
  \frac{\hbar^2}{2m} \left( \frac{4c^2+3-8c}{4u^2} + \frac{u^2 k^4}{16} - \frac{ck^2}{2} + k^2 a \right) = V-E.
 \ee
Eq.~(\ref{37}) can be split up into two parts. The first part for the potential as 
  \bea
 \label{38}
  V(u)&=&\frac{\hbar^2}{2m} \left( \frac{4c^2+3-8c}{4u^2} + \frac{u^2 k^4}{16}  \right),\\\nonumber
  &=&A u^2+\frac{B}{u^2}
 \eea
 where $A= \frac{k^4}{16}$ , $B=\frac{4c^2+3-8c}{4}$. In terms of the variable $x$, we have
    \be
  \label{312}
V(x) = A x^2 + \frac{B}{x^2}.
 \ee
 The second part is
   \be
   \label{390}
 E=\frac{\hbar^2 k^2}{4m} \left( c-2a \right).
 \ee
For the three-dimensional harmonic oscillator with energy dependent potential of the form $V(r,E)=\frac{m \omega^2 f(E_n) r^2}{2}$, the Schr{\"o}dinger equation reads
   \be
  \label{313}
\frac{d^2 \phi}{dr^2} + \frac{2m}{\hbar^2} \left[ E- \frac{m \omega^2 f(E_n) r^2}{2} - \frac{\ell (\ell +1) \hbar^2}{2 m r^2 }\right] \phi = 0,
\ee
where as usual $V_{eff} (r)=\frac{m \omega^2 f(E_n) r^2}{2} +  \frac{\ell (\ell +1) \hbar^2}{2 m r^2 }$ is the effective potential. The solutions of Eq.(\ref{313}) are  
\be
  \label{310}
\phi(u) = u^{\frac{2c-1}{2}} e^{\frac{- u^2 k^2}{8}}  {}_1 F_1\left(a; c ; \frac{u^2 k^2}{4} \right).
 \ee
Comparing $V_{eff} (r)$ with Eq.(\ref{38}), we obtain 
   \be
  \label{314}
k^2=  \frac{4 m \omega}{\hbar} f(E_n)^{\frac{1}{2}},
\ee
  \be
  \label{315}
c=\frac{1}{2} (1-2 \ell), c= \frac{1}{2} (3+2\ell), \quad  c \in \mathbb{Z^-}.
\ee
In the third step, we study the saturation of energy for harmonic oscillator using 
  \be
  \label{316}
E_n = f(E_n)^{\frac{1}{2}} E_n^{(HO)},
\ee
where  $E_n^{(HO)}=\hbar \omega \left( n+\frac{3}{2} \right)$. Then, using Eq.(\ref{390}), we have
  \be
  \label{317}
\hbar \omega \left( n+\frac{3}{2} \right)f(E_n)^{\frac{1}{2}} = \frac{\hbar^2 k^2}{4m} (c-2a).
\ee
Substituting for $c$ and $k^2$ from Eqs.~(\ref{314}-\ref{315}), we obtain
  \be
  \label{319}
 \left( n+\frac{3}{2} \right) = \frac{1}{2} (3+2\ell) - 2a.
\ee
from which we deduce that $a= \frac{\ell-n}{2}$. Using Eq.(\ref{310}) for $u=r$, the eigenfunctions read 
 \be
  \label{320}
  \phi_{n \ell}(r) = C_{n \ell}  r^{\ell +1}  e^{\frac{- r^2 m \omega f(E_n )^{\frac{1}{2}}}{2\hbar}}  {}_1 F_1\left(\frac{\ell - n }{2}; \ell + \frac{3}{2} ; \frac{r^2 k^2}{4} \right).
\ee
In the fourth step, we should provide the functional form of $f(E_n)$. By proposing that $f(E_n )= (1+ \lambda En)^q$, the energy can be calculated for different cases depending on the different values of $q$, using Eq. (\ref{316}), as
  \be
  \label{321}
 E_n =  (1+ \lambda En)^{\frac{q}{2}} E^{(HO)}_n.
\ee
For $q=0$
  \be
  \label{322}
 E_n = E^{(HO)}_n.
\ee
For $q=1$
  \be
  \label{323}
 E_n = E^{(HO)}_n \frac{\lambda E^{(HO)}_n \pm \sqrt{\lambda^2 ( E^{(HO)}_n)^2 +4}}{2}.
\ee
For $q=2$
  \be
  \label{324}
 E_n = \frac{E^{(HO)}_n}{1- \lambda E^{(HO)}_n}.
\ee
For $q=4$
  \be
  \label{325}
 E_n =\frac{1-2 \lambda E^{(HO)}_n \pm \sqrt{1-4\lambda  E^{(HO)}_n} }{2 \lambda^2  E^{(HO)}_n}.
\ee
\\
We next discuss the results for different cases. In Figures~(\ref{H1}-\ref{H4}), we plotted the energy as a function of the quantum number $n$ at different value of $\lambda$ and for different values $q$. The saturation in this system
could be approached with negative choices of $\lambda$ and more noticeably with decreasing (toward $-\infty$). For positive choices of $\lambda$, the saturation could not be reached with physically meaningful results. Also, the presence of the exponent $q$ makes the quenching of the curves faster and thus more profound saturation effect. According to result Eq.~(\ref{325}), as the quantum number $n\to\infty$, the energy $E^{(HO)}\to\infty$, and the energy $E_n$ becomes saturated to the value $-\frac{1}{\lambda}$:
\be
\label{sat1}
\lim_{n\to\infty} E_{n}=-\frac{1}{\lambda},
\ee
This coincides with the results found in other potential models; see for instance \cite{al2018saturation}.


 \newpage
 \subsection{The hydrogen Atom}
\begin{figure}[!htbp]
  \centering
  \subfigure [] {\includegraphics[scale=0.7]{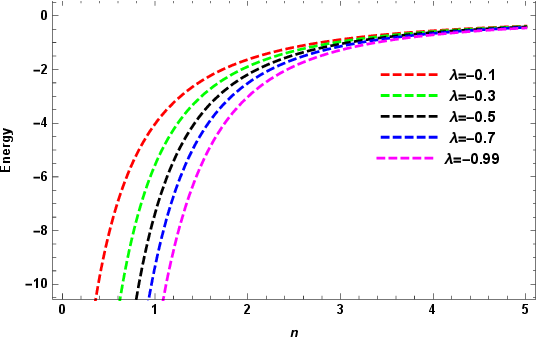}}\quad
  \subfigure []{\includegraphics[scale=0.7]{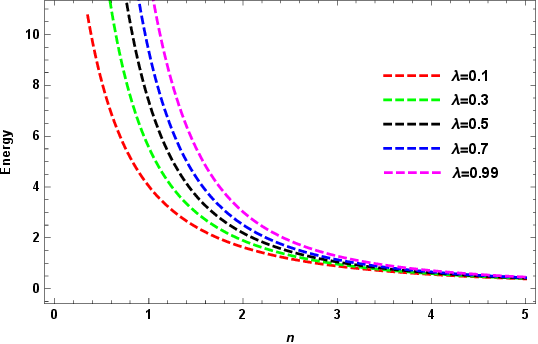}}
  \hfill
    \caption{ $E_n$ as function of $n$ for the linear E-dependent Hydrogen atom, with $q=1$ and different values of $\lambda$ in (GeV$^{-1}$).}
	\label{Ha3}
\end{figure}

\begin{figure}[!htbp]
  \centering
  \subfigure [] {\includegraphics[scale=0.7]{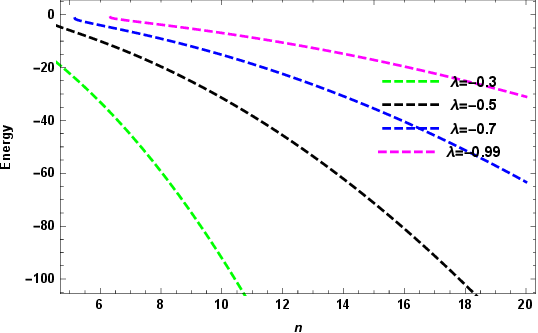}}\quad
  \subfigure []{\includegraphics[scale=0.7]{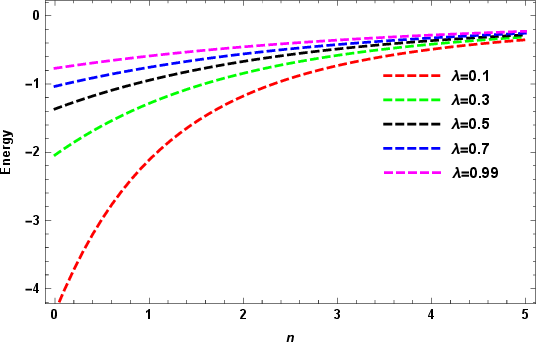}}
  \hfill
   \caption{ $E_n$ as function of $n$ for the linear E-dependent Hydrogen atom, with $q=4$ and different values of $\lambda$ in (GeV$^{-1}$).}
	\label{Ha1}
\end{figure}

The Hamiltonian for a particle in a Coulomb field (hydrogen atom) is:
\be
  \label{326}
 H= -\frac{\hbar^2}{2m} \left(\frac{d^2}{dr^2} + \frac{2}{r} \frac{d}{dr} + \frac{\ell(\ell+1)}{r^2} \right) - \frac{a}{r},
\ee
where $a=\frac{e^2}{4 \pi \epsilon_0 }$. In the ground state, $\ell=0$ and the Hamiltonian becomes 
\be
  \label{327}
 H= -\frac{\hbar^2}{2m} \left(\frac{d^2}{dr^2} + \frac{2}{r} \frac{d}{dr} - \frac{a}{r}\right).
\ee
We calculate the energy spectrum of the Hydrogen atom using the confluent hypergeometric equation. Energy spectrum $E_n$, and solutions $\phi(u)$ for several energy-dependent potentials $V (E_n; u)$ obtained from Sec.2 when it is applied to the confluent hypergeometric equation of the Hydrogen atom.
For the hydrogen atom, the choice \cite{garcia2009exactly}
\be
  \label{328}
 g(x)= \frac{k^2}{x}
\ee
can generate the eigen-solutions. Using Eq.(\ref{215}), and the coordinate change in Eq. (\ref{212}), we have 
\be
  \label{329}
u = \pm \int^x \frac{dx}{\sqrt{\frac{k^2}{x} x}} = \pm \frac{x}{k}.
\ee
Changing the variable  $u= \pm \frac{x}{k}$ then yields 
\be
  \label{330}
 F(u)= \pm u k.
\ee
Using Eq. (\ref{218}) with $F^{''} (u)= 0$,  $Q(F(u))=(c-u k)$, we obtain $ W(u)$ as
\be
  \label{331}
 W(u)= \frac{c}{2u} - \frac{k}{2}.
\ee
Substituting this result in Eq.~(\ref{223}), we have
\be
  \label{332}
 v(u)= \frac{c^2}{4u^2} + \frac{k^2}{4}-\frac{c}{2u}k-\frac{c}{2u^2}.
\ee
Thus, using Eq. (\ref{224}), we obtain
\be
  \label{333}
\frac{\hbar^2}{2m} \left( \frac{c^2-2c}{4u^2} -\left( \frac{c k^2}{2uk} + \frac{a k^2}{uk}\right)  \right)=V-E.
\ee
The following separation of Eq.(\ref{333}) into two parts is possible
\be
  \label{334}
\frac{\hbar^2}{2m} \left( \frac{c(c-2)}{4u^2} +\frac{(2a-c) k}{2u} \right)=V,
\ee
\be
  \label{335}
-\frac{\hbar^2 k^2}{8m} =E.
\ee
In the second step, we use the solution of the Schr{\"o}dinger equation obtained in Eq. (\ref{221}), we obtain 
\be
  \label{336}
\phi (u) =   u^{\ell -1}  e^{\frac{- u   f(u )}{n a_{0}}}  {}_1 F_1\left(\ell+1 - n ; 2(\ell + 1) ; \frac{2f(E_n )u}{n a_0} \right).
\ee
We also calculate the energy for different cases depending on the different values of $q$ as 
\be
  \label{337}
E_n = (1+\lambda E_n)^{\frac{q}{2}} E_n^{(H)},
\ee
where we have chosen $f(E_n )=(1+\lambda E_n)^q$.
For $q=0$
  \be
  \label{322}
 E_n = E^{(HO)}_n.
\ee
For $q=1$
  \be
  \label{323}
 E_n = E^{(HO)}_n \frac{\lambda E^{(HO)}_n \pm \sqrt{\lambda^2 ( E^{(HO)}_n)^2 +4}}{2}.
\ee
For $q=2$
  \be
  \label{324}
 E_n = \frac{E^{(HO)}_n}{1- \lambda E^{(HO)}_n}.
\ee
For $q=4$
  \be
  \label{325}
 E_n =\frac{1-2 \lambda E^{(HO)}_n \pm \sqrt{1-4\lambda  E^{(HO)}_n} }{2 \lambda^2  E^{(HO)}_n}.
\ee
We next discuss the results for different cases. In Figures~(\ref{Ha1}-\ref{Ha3}), we plotted the energy as a function of the quantum number $n$ at different values of $\lambda$ and for different values $q$. The saturation in this system
is reached with negative choices of $\lambda$ and faster with increasing its negativity. Also, the presence of the exponent $q$ makes the saturation faster to be achieved. According to result Eq.~(\ref{325}), as the quantum number $n\to\infty$, the energy $E^{(HO)}\to\infty$, and the energy $E_n$ becomes saturated to the value $-\frac{1}{\lambda}$:
\be
\label{sat1}
\lim_{n\to\infty} E_{n}=-\frac{1}{\lambda},
\ee
which also coincides with the results in other potential models; see for instance \cite{al2018saturation}.

\begin{figure}[!htbp]
  \centering
  \subfigure [] {\includegraphics[scale=0.7]{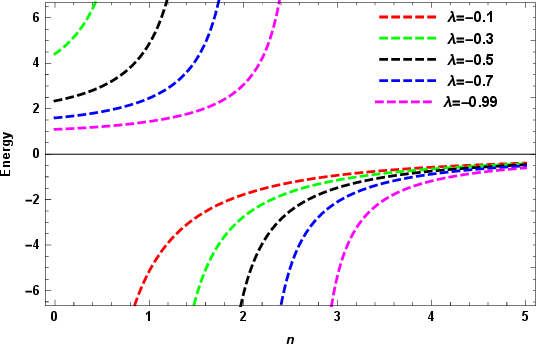}}\quad
  \subfigure []{\includegraphics[scale=0.7]{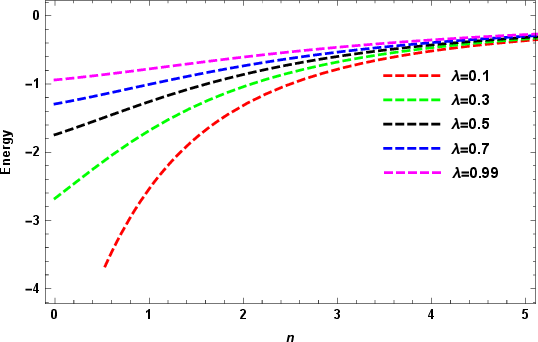}}
  \hfill
  \caption{ $E_n$ as function of $n$ for the linear E-dependent Hydrogen atom, with $q=2$ and different values of $\lambda$ in (GeV$^{-1}$).}
	\label{Ha2}
\end{figure}

\newpage

\subsection{Heavy quarkonia}
\begin{figure}[!htbp]
  \centering
  \subfigure [] {\includegraphics[scale=0.7]{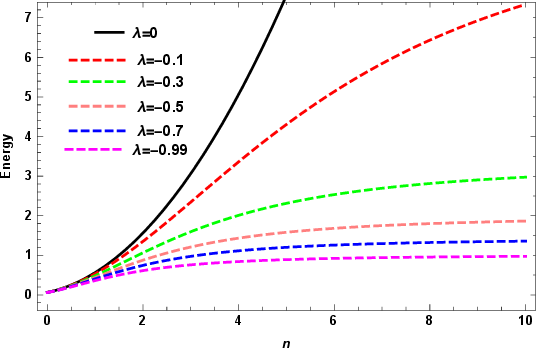}}\quad
  \subfigure []{\includegraphics[scale=0.7]{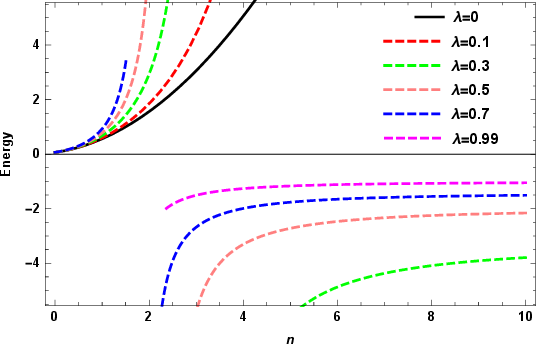}}
  \hfill
    \caption{ $E_n$ as function of n for the Eckart potential, with  $q=1$ and different values of $\lambda$ in (GeV$^{-1}$).}
	\label{fig:f38}
\end{figure}

In this section, we study the mass of the spectrum, the potentials of heavy quarks dependent on energy through the method, and apply it to the hypergeometric equation.
According to Table 3. of \cite{garcia2009exactly}, the hypergeometric equation, not the confluent, with suitable choice of $g(x)$ can generate potentials that are appropriate for confined two-particle quantum systems such as heavy quarkonia. The hypergeometric equation is
\be
\label{41}
x(1-x)y^{''}+ [c-(a+b+1)x ]y^{'}-aby=0.
\ee
Its first solution is:
\be
\label{42}
y(x)={}_2 F_1\left(a,b; c;x \right).
\ee
For our purposes, we select the third case of the energy spectrum presented in Table 3. of Ref \cite{garcia2009exactly}, viz., 
\be
\label{43}
E_n= - \frac{k^2}{16} \left[ (2n+1)^2 + \frac{p^2}{(2n+1)^2}\right],~~~n=0,1,2,...
\ee
This result is in agreement with the result reported in \cite{doi:10.1142/S0217732319503073} where the trigonometric Rosen-Morse potential for confined systems is reduced to Nikiforov-Uvarov problem for solving hypergeometric differential equations. If the change $k^2 \to k^2  f(\epsilon_n)$ is realized, the potential has an energy-dependent form. Suppose: 
\be
\label{44}
f(\epsilon_n)= (1+ \lambda E_n)^q,
\ee
we then have
\be
\label{45}
E_n= - \frac{k^2 (1+ \lambda E_n)^q}{16} \left[ (2n+1)^2 + \frac{p^2}{(2n+1)^2}\right].
\ee
To tidy up the notations, we define
\be
\label{46}
\beta \equiv - k^2\left[ (2n+1)^2 + \frac{p^2}{(2n+1)^2}\right],
\ee
then we obtain
\be
\label{47}
E_n=  \frac{(1+ \lambda E_n)^q}{16}\beta.
\ee
The assignment of numerical value for $q$ determines the next steps. As an application, we study heavy quarkonia mass spectra with $q=1$. The energy spectrum from Eq. (\ref{44}) is then
\be
\label{48}
E_n=  \frac{\beta}{16- \lambda \beta}.
\ee
Rewriting Eq.~(\ref{48}) as $E_n=\frac{1}{\frac{16}{\beta}- \lambda}$ and taking the limit $n\rightarrow\infty$, which by Eq.~(\ref{46}) means that $\beta\rightarrow\infty$, we arrive at
\be
\label{sat1}
\lim_{n\to\infty} E_{n}=-\frac{1}{\lambda},
\ee
as expected. We next discuss the results for different cases. In Figure~(\ref{fig:f38}),  the energy is plotted as a function of the quantum number $n$ at different value of $\lambda$ and for $q=1$. The saturation in this system behaves the same as in the hydrogen system and the harmonic oscillator as well.
\\
As an application, we investigate the saturation effect in heavy quarkonia mass spectra. This system of bound states is of great important tool in particle physics. Heavy quarkonia systems have a spin-averaged mass related to the energy via the relation \cite{al2018saturation}:
\be
\label{49}
M_{q \Bar{q}}=m_q+m_{\Bar{ q}}  +E_{n \ell},
\ee
where $E_{n \ell}$ is given from Eq.~(\ref{46}), and $m_q$ is the mass of the constituent quark. We adopt the values $m_c =1.209$ GeV and $m_b =4.350$ GeV. The free parameters $k$ and $p$ appeared in Eq.~(\ref{46}) were adjusted by fitting simultaneously, for each value of $\lambda$, the mass splittings $M_{2S}-M_{1S}$ and $M_{4S}-M_{3S}$ for $(c\bar{c})$, and $M_{3S}-M_{1S}$ and $M_{4S}-M_{2S}$ for $(b\bar{b})$. These choices were adopted in this work as they were found to generate more accurate values for the mass spectra. Also, using the idea of the splitting reduces the dependency on the initial choice of the mass of the quarks $m_q$ as can be seen from the Eq~(\ref{49}):
\be
M_{2S}-M_{1S}= E_{2S}-E_{1S},
\ee
and so on. With these setups, we found different values for the parameters $k$ and $p$ for each bound state system and for each value of the saturation parameter $\lambda$. After careful choices for the quark masses $m_q$ to obtain the best fit, we adopted the values $m_c=1.697$ GeV and $m_b=4.568$ GeV, and we then generated the spectral mass values of $(c\bar{c})$ and $(b\bar{b})$ for  $(\ell = 0)$ and ($n=0,1,2,3,4,\cdots$) using the Eqs. (\ref{48}) and (\ref{45}) at two different values of $\lambda$. The results are summarized in Table \ref{tab:4-4} and \ref{tab:4-5} along with the corresponding available experimental values \cite{Workman:2022ynf}. According to these results, it is found that the choice $\lambda=-0.4$ yields results that are in better agreement with the experimental data than these for the choice  $\lambda=-0.2$ for the charmonium, while the choice $\lambda=-0.6$ yields in favor of $\lambda=-0.2$ for the bottomonium. It is also clear that both bound state systems are in favor of including the energy-dependence effect as it produces saturated mass spectra and in better agreement with experimental data.

\begin{table}[!htbp] 
\centering 
\caption{The produced spin-averaged mass spectra of $c\bar{c}$ in GeV at different values of $\lambda$ in GeV$^{-1}$. The value $m_c=1.697$ GeV is adopted. Experimental data are taken from \cite{Workman:2022ynf}.} 
\label{tab:4-4}
\begin{tabular}{cccccc}
\hline
$nL$ & $\lambda=0$& $\lambda=-0.2$ & $\lambda=-0.4$&Experimental\\ \hline  
$1S$ & 3.010  & 2.973  & 3.097 & $ 3.096$\\
$2S$ & 3.563 & 3.526 & 3.650 & $3.649$\\
$3S$ & 3.841 & 3.838 & 4.041 & $4.040$\\
$4S$ & 4.216 & 4.213 & 4.416 & $4.415$\\
$5S$ & 4.707 & 4.622 & 4.729 &---\\
$6S$ & 5.318 & 5.032 & 4.972 & ---\\
$\vdots$ & $\vdots$ & $\vdots$ & $\vdots$ & $\vdots$\\
$9S$ & 7.879 & 6.085 & 5.403 & ---\\
\hline
\end{tabular}
\end{table}
\begin{table}[!htbp] 
\centering 
\caption{The produced spin-averaged mass spectra of $b\bar{ b}$ in GeV at different values of $\lambda$ in GeV$^{-1}$.  The value $m_b=4.568$ GeV is adopted. Experimental data are taken from \cite{Workman:2022ynf}.} 
\label{tab:4-5}
\begin{tabular}{cccccc}
\hline
$nL$ & $\lambda=0$& $\lambda=-0.3$ & $\lambda=-0.6$&Experimental\\ \hline  
$1S$ & 8.500  & 8.878  & 9.471 & $ 9.460$\\
$2S$ & 9.151 & 9.228 & 10.002 & $10.023$\\
$3S$ & 9.395 & 9.489 & 10.383 & $10.355$\\
$4S$ & 9.707 & 9.784 & 10.558 & $10.580$\\
$5S$ & 10.112 & 10.092 & 10.645 &$10.579$\\
$6S$ & 12.716 & 10.389 & 10.694 & ---\\
$\vdots$ & $\vdots$ & $\vdots$ & $\vdots$ & $\vdots$\\
$9S$ & 12.716 & 11.103 & 10.753 & ---\\
\hline
\end{tabular}
\end{table}
\newpage
\section{Summary and conclusion}
In this work, we investigated the saturation effect in the energy or mass spectra of three quantum models with energy-dependent potentials: the harmonic oscillator, the hydrogen atom, and the heavy quarkonia. These potential models were generated using the method proposed in \cite{garcia2009exactly}, which is based on studying various canonical point and gauge transformations applied to a function, called $g(x)$, multiplied by a given differential equation of known solutions as special orthogonal functions, which convert it into a Schr{\"o}dinger-like equation. The first two models are produced by implementing the method on the confluent hypergeometric differential of the well-known solutions ${}_1 F_1$, while the heavy quarkonia potential model is produced from the hypergeometric differential of the well-known solutions ${}_2 F_1$. In all models, with energy dependence factor of the form $f(E_n)= (1+ \lambda E_n)^q$, we have found that the saturation can be reached for negative values of the parameter $\lambda$ for some values of $q$, and that in the limit of infinite quantum number $n\rightarrow\infty$, the energy tends to saturate to the value proportional to $\frac{-1}{\lambda}$.  In particular, the heavy quarkonia mass spectra for both $c\bar c$ and $b\bar b$ were produced at different values of $\lambda$ with the choice $q=1$. The mass spectra were compared with the available experimental data, and found that the choices $\lambda=-0.4$ and $\lambda=-0.6$ produce the mass spectra for charmonium and bottomonium, respectively, with good agreement with the available experimental data.
\bibliography{ref}
\bibliographystyle{unsrt}
\end{document}